# Barriers and Enablers of Online Instruction in Hospitality Education in the Philippines: An Exploratory Study

Maria Anna D. Cruz[1], Jeaneth D. Serna[1], Lloyd D. Feliciano[1], Mike Haizon M. David[1], Ma. Ferna Bel L. Punsalan[1], Glen Brian L. Lacsa[1], Michelle C. Castro[1], John Paul P. Miranda[1*]

1. **Pampanga State University**, Pampanga, Philippines

* **Correspondence:**
John Paul P. Miranda, Pampanga State University, jppmiranda@pampangastateu.edu.ph

## ABSTRACT

This study examined the barriers and enablers of online instruction in hospitality education. A sequential exploratory design was implemented with hospitality teachers from both public and private higher educational institutions in the Philippines. Thematic analysis of interviews identified four key themes: technological barriers, pedagogical challenges, institutional and personal support, and integration of artificial intelligence (AI). These themes were transformed into survey constructs and tested for reliability. Pedagogical challenges, including difficulties in teaching hands-on subjects and sustaining student engagement, emerged as the most critical concerns. Technological barriers such as unstable internet and limited devices were moderately rated, while institutional and personal support received mixed evaluations. Teachers viewed AI integration as helpful but also expressed caution and emphasized the need for training. Reliability analysis showed acceptable to good internal consistency across constructs. The findings highlight the importance of strengthening pedagogical training, providing clear institutional support, and fostering responsible competence in AI use. Future studies should validate these results with larger and more diverse samples.

# INTRODUCTION

Hospitality education relies heavily on experiential-, simulation-, and service-based learning (de la Mora Velasco et al., 2025; Lin et al., 2017; Melo et al., 2022). In the Philippines, where tourism and hospitality contributed about 8.6% of the country's GDP in 2023 and generated ₱2.09 trillion in direct gross value added (Philippine Statistics Authority, 2024), online instruction presents significant challenges. Teachers often face unstable internet connectivity, limited facilities, and difficulties in replicating hands-on practice in virtual environments (Atingo & Uluma, 2020; Chandra et al., 2022; Kamble & Dandge, 2025; Stylianou & Pericleous, 2025). This is true in the case of the Philippines (Arante & Bascon, 2025; Baticulon et al., 2021; Fabito et al., 2021) where such barriers affects student engagement (Castulo et al., 2025; Montejo et al., 2025), the quality of feedback (Castulo et al., 2025; Galon, 2024), and the attainment of professional competencies that are central to hospitality education (Miranda & Cruz, 2023).

Existing studies have highlighted student satisfaction, readiness, and digital skills in online environments (Deale et al., 2021; Goh & Sigala, 2020), but fewer have focused on the perspectives of hospitality teachers, particularly in the Philippine context. Teachers play a central role in shaping learning strategies, evaluating performance, and ensuring that practical competencies are achieved (de la Mora Velasco et al., 2025; Dominguez Picart et al., 2025; Ochoa, 2022). Their perspective is therefore essential in identifying both barriers that hinder instruction and enablers that make online teaching feasible (Coman et al., 2020; Gkrimpizi et al., 2023). While technological innovations such as artificial intelligence are beginning to appear in educational practice, the more urgent question remains how teachers themselves navigate the broader challenges of online instruction.

International research on online education identifies digital readiness, learner engagement, and teacher adaptability as recurring areas of concern. (Coman et al., 2020) described how these factors influence academic performance during emergency remote teaching. (Stylianou & Pericleous, 2025) reported that graduates of tourism and hospitality programs require strong digital skills to succeed in the post-pandemic environment. These studies originate mainly from developed regions where infrastructure and institutional support are stable. The Philippine context differs because access to reliable internet, training, and administrative systems remains uneven (Cynthia et al., 2023; Espinosa et al., 2025). This contrast establishes the rationale for examining how hospitality educators in the Philippines manage online instruction under constrained conditions.

This study addresses this gap. It seeks to answer the following questions: (1) What are the barriers and enablers of online instruction in hospitality education? (2) How do these barriers and enablers form constructs derived from teacher responses? and (3) How do hospitality teachers rate these constructs in terms of agreement, and are the constructs internally consistent?

# METHODOLOGY

## Research Design

This study used a sequential exploratory design that combined qualitative and quantitative procedures. The qualitative phase identified the categories of barriers and enabling conditions in online instruction. The quantitative phase developed a survey from these categories and tested internal consistency. The process included four steps: (1) qualitative data collection through interviews, (2) thematic analysis and category development, (3) survey construction based on verified themes, and (4) quantitative analysis of teacher responses. Figure 1 presents the sequence of these stages.

**Figure 1.**
*Sequential exploratory design of the study*

**Cruz et al. (2026)**

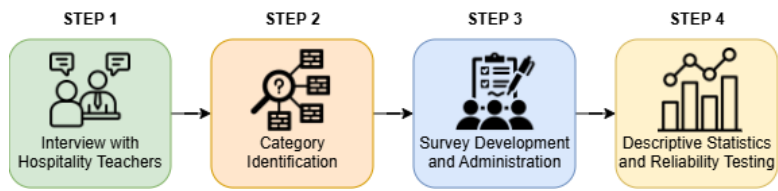

## Research Locale, Sample, and Sampling Design

The study was conducted among hospitality teachers in higher education institutions in the Philippines. A total of sixty-eight (68) teachers participated, including thirty-eight (38) from public institutions and thirty (30) from private institutions. The participants were selected through purposive sampling, which allowed the inclusion of teachers with actual experience in online or hybrid instruction during and after the pandemic. Invitations were sent through institutional email lists and professional networks. Participation was voluntary. Most of the participants were instructors (n = 48), while the rest were associate professors (n = 7), assistant professors (n = 7), and professors (n = 6). In terms of employment status, forty-four (44) were non-permanent and twenty-four (24) were permanent. Teaching experience ranged from one (1) to thirty-two (32) years. The largest group had one to three years of experience (n = 27), while sixteen (16) reported four to ten years, and seventeen (17) had more than fifteen years of experience. For online instruction, most had taught four semesters (n = 21), followed by two semesters (n = 13) and six semesters (n = 11). Only a few had more than ten semesters of online teaching.

## Research Instrument and Data Gathering Procedure

Two phases of data collection were conducted. The first phase involved interviews to identify the barriers and enablers of online instruction. Both in-person and online interviews were conducted. In-person interviews were done with teachers who were accessible to the researchers. Online interviews were arranged for teachers who were located far from the research team. The interviews asked about challenges in online teaching, strategies that helped, and experiences with AI tools.

The second phase involved the administration of a survey instrument. The survey items were developed from the themes that emerged in the first phase. The constructs were not predetermined at the start. They were identified after coding and grouping the qualitative data. The survey consisted of Likert-scale items rated from 1 (Strongly Disagree) to 5 (Strongly Agree). Both in-person and online surveys were administered depending on the accessibility of the respondents.

## Data Gathering and Analysis

The thematic coding procedure followed a structured sequence. First, transcripts were read several times to achieve complete familiarity with the content. Second, key phrases that represented barriers or supports were labeled as codes. Third, codes that expressed similar ideas were grouped into subcategories. Fourth, broader categories were defined as main themes. Two researchers coded independently, and two others reviewed disagreements until a single interpretation was approved. Figure 2 shows the process applied during thematic coding.

The quantitative data were analyzed using descriptive statistics. Means and standard deviations were computed for each item to determine the level of agreement among teachers. Cronbach's alpha was calculated for each construct to test internal consistency.

**Figure 2.**
*Thematic coding procedure*

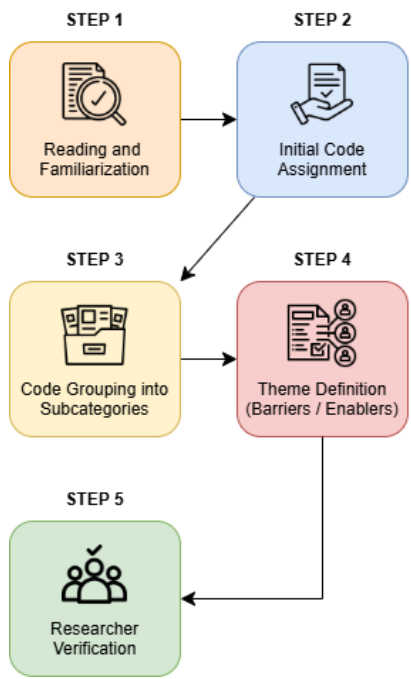

## INITIAL RESULTS

Table 1 presents the emergent barriers and enablers of online instruction in hospitality education. Four major themes were identified: technological barriers, pedagogical challenges, institutional and personal support, and AI integration in teaching. Technological barriers were raised by teachers who cited unstable internet, lack of devices, and weak infrastructure. Nine teachers shared, "Students do not have stable internet connection at home," while six noted the absence of proper equipment such as laptops and microphones. Issues with learning management systems and power interruptions were also mentioned, as one teacher stated, "Connectivity issues. Some of the students have bad internet connections that disrupt their learning," and another added, "Power interruptions often disrupt my classes."

Pedagogical challenges were also prominent. Teachers emphasized the difficulty of teaching hands-on subjects, maintaining student engagement, and providing feedback. Ten teachers said, "The actual laboratories," while eight explained, "Hands-on activities like cooking and baking require close supervision." Twelve respondents noted, "The active participation of the learners," and nine stated, "Students do not participate in class." Similar concerns were echoed in feedback and assessment, as one respondent said, "Hands-on assessment is hard," while another shared, "Giving performance-based feedback online is challenging."

The results correspond with studies that examined the limits of online instruction for subjects that depend on demonstration and performance. (Melo et al., 2022) observed that simulation can support practical learning but cannot reproduce the supervision present in physical laboratories. (de la Mora Velasco et al., 2025) also reported that hospitality teachers continue to rely on physical settings to develop applied competencies. The responses in the present study indicate that some teachers have begun to use artificial-intelligence tools for lesson preparation and student practice. This development suggests a gradual movement toward blended formats that combine online delivery with limited in-person application.

Institutional and personal support was highlighted by teachers who acknowledged limited training, unclear guidelines, and increased workload. Seven reported, "Our school provides training but support is sometimes limited," while others mentioned, "Online teaching increases my workload and leads to burnout."

**Cruz et al. (2026)**

Research on digital transformation in higher education emphasizes the importance of organizational readiness and administrative clarity. (Gkrimpizi et al., 2023) concluded that institutional barriers often originate from undefined policies and inconsistent technical support. The present data illustrate a similar condition in Philippine institutions, where limited guidance and unequal access to training reduce the capacity of teachers to innovate within online classrooms.

AI integration was viewed as both an enabler and a risk. Teachers expressed appreciation for its applications, such as, "ChatGPT helped me improve my online teaching" and "AI provides accurate information," while others raised concerns, cautioning to "Carefully use the AI generated answers." Some also emphasized the need for training, with one teacher stating, "I recommend… training educators on the effective use of AI."

The mixed perspectives on artificial intelligence reflect concerns discussed in international literature. (Goh & Sigala, 2020) explained that educators value the efficiency of digital tools but remain cautious about accuracy, originality, and ethical use. The teachers in this study expressed comparable views and requested structured training before AI is fully integrated into instruction.

**Table 1.**
*Emergent barriers and enablers of online instruction in hospitality education*

| Major Themes | Sub-theme | Sample Responses (frequency) |
|---|---|---|
| Technological Barriers | Connectivity issues | "Students do not have stable internet connection at home" (n=9) |
| | Device and equipment gaps | "Internet connection and equipment like laptop and condenser mic" (n=6) |
| | System and infrastructure | "Connectivity issues. Some of the students have bad internet connections that disrupt their learning." (n=7) |
| | | "Power interruptions often disrupt my classes" (n=5) |
| Pedagogical Challenges | Hands-on instruction | "The actual laboratories" (n=10) |
| | | "Hands-on activities like cooking and baking require close supervision…" (n=8) |
| | Student engagement | "The active participation of the learners" (n=12) |
| | | "Students do not participate in class" (n=9) |
| | Feedback and assessment | "Hands-on assessment is hard" (n=6) |
| | | "Giving performance-based feedback online is challenging" (n=5) |
| Institutional and Personal Support | Training and guidance | "Our school provides training but support is sometimes limited." (n=7) |
| | Admin/technical support | "Clear guidelines do not always exist for using digital tools." (n=5) |
| | Workload and wellbeing | "Online teaching increases my workload and leads to burnout." (n=9) |
| | | "Balancing online teaching with family responsibilities is difficult." (n=6) |
| AI Integration in Teaching | Positive applications | "ChatGPT helped me improve my online teaching" (n=14) |
| | | "AI provides accurate information." (n=11) |
| | | "AI tools have positively influenced my teaching by helping me create lesson plans…" (n=10) |
| | Concerns and risks | "Carefully use the AI generated answers." (n=8) |

| Major Themes | Sub-theme | Sample Responses (frequency) |
|---|---|---|
| | Need for training | "I recommend… training educators on the effective use of AI." (n=6) |

The survey items derived from the interviews were administered to the same teachers. Reliability analysis showed that three constructs had good internal consistency: technological barriers (α = 0.807), pedagogical challenges (α = 0.819), and AI integration in teaching (α = 0.810). Institutional and personal support had acceptable reliability (α = 0.724).

As shown in Table 2, pedagogical challenges had the highest means. Teachers agreed that student misuse of AI tools (M = 4.44, SD = 0.74), adjusting strategies (M = 4.40, SD = 0.72), and teaching hands-on subjects (M = 4.35, SD = 1.02) were the most pressing concerns. Feedback (M = 4.06, SD = 0.96) and reminders to complete tasks (M = 4.07, SD = 1.01) also reflected strong agreement. Technological barriers were moderately rated, with unstable internet (M = 3.84, SD = 1.03) and solving problems alone (M = 3.75, SD = 1.04) scoring higher than noisy teaching spaces (M = 2.43, SD = 1.19) and lack of private space (M = 2.74, SD = 1.28).

Institutional and personal support received mixed evaluations. Encouragement from departments (M = 4.09, SD = 0.89) scored higher than training (M = 3.66, SD = 0.96) and administrative support (M = 3.66, SD = 1.02). Items related to workload (M = 3.24, SD = 1.13), burnout (M = 3.18, SD = 1.17), and balancing family responsibilities (M = 3.15, SD = 1.12) were moderately rated. AI integration was viewed favorably, with awareness of AI tools (M = 4.12, SD = 0.72) and their usefulness for saving time (M = 3.90, SD = 0.95) rated highly, while actual use for content generation (M = 3.60, SD = 1.08) was lower. Teachers also recognized the need for additional AI training (M = 3.85, SD = 0.80).

**Table 2.**
*Descriptive statistics*

| Construct | Item | Mean | SD |
|---|---|---|---|
| Technological Barriers | I experience challenges because the internet is unstable. | 3.84 | 1.03 |
| | I experience challenges because I have limited access to reliable devices. | 2.85 | 1.08 |
| | I experience challenges because the LMS often has technical issues. | 3.4 | 1.13 |
| | I experience challenges because I must solve technical problems alone. | 3.75 | 1.04 |
| | I experience challenges because I teach in a noisy space at home. | 2.43 | 1.19 |
| | I experience challenges because power interruptions disrupt my classes. | 3.65 | 1.31 |
| | I experience challenges because I multitask between teaching and duties. | 3.13 | 1.37 |
| | I experience challenges because I lack a private teaching space. | 2.74 | 1.28 |
| Pedagogical Challenges | I find it difficult to teach hands-on subjects online. | 4.35 | 1.02 |
| | I find it difficult to help students apply concepts. | 4.12 | 0.82 |
| | I find it difficult to give performance-based feedback. | 4.06 | 0.96 |
| | I find it difficult to adjust my teaching strategies. | 4.4 | 0.72 |
| | I find it difficult to encourage student participation. | 4.03 | 1.08 |
| | I find it difficult to address passive learning. | 3.71 | 1.13 |
| | I find it difficult to ensure students complete tasks. | 4.07 | 1.01 |

| Construct | Item | Mean | SD |
|---|---|---|---|
| | I find it difficult to prevent students from misusing AI tools. | 4.44 | 0.74 |
| Institutional and Personal Support | I receive adequate support when my school provides training. | 3.66 | 0.96 |
| | I receive adequate support when admin and technical help is available. | 3.66 | 1.02 |
| | I receive adequate support when clear guidelines exist. | 3.43 | 1.04 |
| | I receive adequate support when my department encourages innovation. | 4.09 | 0.89 |
| | I feel that online teaching increases my workload. | 3.24 | 1.13 |
| | I feel burned out after online classes. | 3.18 | 1.17 |
| | I feel it is difficult to balance teaching and family responsibilities. | 3.15 | 1.12 |
| | I feel confident teaching in online or hybrid settings. | 3.84 | 0.94 |
| AI Integration in Teaching | I am aware of AI tools for teaching. | 4.12 | 0.72 |
| | I use AI tools to generate content or quizzes. | 3.6 | 1.08 |
| | I find AI tools helpful in saving time. | 3.9 | 0.95 |
| | I need more training to use AI tools effectively. | 3.85 | 0.8 |

## CONCLUSION AND FUTURE RESEARCH

The study identified the barriers and enablers of online instruction in hospitality education. Four themes were generated from the interviews and validated through a survey. Results showed that pedagogical challenges such as teaching hands-on subjects and encouraging participation were the most critical concerns, while institutional and personal support received mixed evaluations. Reliability analysis confirmed acceptable internal consistency. The results imply that hospitality programs should strengthen pedagogical training for online delivery, provide clearer support systems, and build teacher competence in using AI tools for instruction.

The study also identified several conditions that support online instruction. Teachers recognized the importance of institutional training programs, encouragement from academic departments, and the responsible use of AI applications as positive elements that sustain online teaching. These aspects represent the enablers of digital pedagogy within hospitality education. Although these supports varied across institutions, they demonstrated that structured guidance, technical assistance, and teacher adaptability can improve the continuity and quality of online instruction. Strengthening these enabling factors can complement efforts to address existing barriers.

Future research should involve a larger number of hospitality teachers to validate the instrument and conduct confirmatory factor analysis. Studies may also explore how AI tools can be integrated into laboratory courses and how institutional mechanisms can address workload and wellbeing issues faced by hospitality educators.

**Cruz et al. (2026)**